# Analysis of the Impact of Central bank Digital Currency on the Demand for Transactional Currency

Ruimin Song;Tiantian Zhao;Chunhui Zhou

January 11, 2024

Student: Ruimin Song;Tiantian Zhao;Chunhui Zhou 21052202013

**Abstract**

  This paper takes the development of Central bank digital currencies as a perspective, introduces it into the Baumol-Tobin money demand theoretical framework, establishes the transactional money demand model under Central bank Digital Currency, and qualitatively analyzes the influence mechanism of Central bank digital currencies on transactional money demand; meanwhile, quarterly data from 2010-2022 are selected to test the relationship between Central bank digital currencies and transactional money demand through the ARDL model. The long-run equilibrium and short-run dynamics between the demand for Central bank digital currencies and transactional currency are examined by ARDL model. The empirical results show that the issuance and circulation of Central bank digital currencies will reduce the demand for transactional money. Based on the theoretical analysis and empirical test, this paper proposes that China should explore a more effective Currency policy in the context of Central bank digital currencies while promoting the development of Central bank digital currencies in a prudent manner in the future.

# 1 Introducere

With the continuous development and application of blockchain, 5G and other information technologies, the emergence of central bank digital currencies is a trend of the times that adapts to the rapid development of the digital economy era, a product of the highly developed modern commodity economy and the continuous advancement of cryptography technology, and a major innovation in the history of mankind.On January 4, 2022, the DC/EP APP hit the shelves of mobile terminal application platforms, and Central bank digital currencies began to truly into people's daily consumption activity scenarios. As of the end of 2022, the stock of DC/EP in circulation reached 13.61 billion. At present, the expansion of consumption scenarios related to Central bank digital currenciesis still the focus of the promotion of DC/EP, and through the active exploration of DC/EP application scenarios around the world, the degree of integration of DC/EP with the life of each of us will become higher and higher. Currency demand is an important factor that has an impact on China's macro-economy, Central bank digital currencies as the latest form of currency evolution, its issuance and circulation will inevitably affect the relationship between the variables of currency demand, will be characterized by the convenience and security of its transactions on the traditional currency demand alternative, making the traditional theory of currency demand has been a great impact on the reduction of the stability of the demand for money, increase the unpredictability of the demand for money. Predictability. At present, China mainly takes money supply as the intermediary index of monetary policy transmission, which makes the central bank face greater difficulties in formulating reasonable money supply targets, and thus affects the implementation of monetary policy. Therefore, studying the impact of Central bank digital currencies on the demand for money is of great theoretical and practical significance for the central bank to grasp the new changes in the demand for money, to improve the effectiveness of China's monetary policy as well as to help the development of China's digital economy and inclusive finance.

The research approach is one of the potential innovations in this paper. The literature on Central bank digital currencies that is currently available is primarily concerned with the meaning of Central bank digital currencies, the creation of pertinent laws and regulations, and the effects on the macroeconomy. There is less empirical analysis of the research on the effects of Central bank digital currencies on the demand for money as a research object. Second, the Baumol-Tobin model has been refined and the transactional money demand function has been constructed based on the features of digital currency that is Central bank digital currencies.

# 2 Review of literature

With the rapid development of the digital economy and digital technology, research on issues related to Central bank digital currencies has become a hotspot and a focus of attention in the theoretical community, and there is a continuous emergence of relevant literature and research on the connotation of Central bank digital currencies and its impact on the demand for money.

Reviewing the development history of the Central bank digital currencies, scholars have respectively clarified the definition of the Central bank digital currencies from the perspectives of the issuer of the Central bank digital currencies, specific forms, etc. Koning (2017)[1]proposes to categorize the Central bank digital currencies into two manifestations according to whether it is based on a central bank account or not: the Central bank digital currencies (CBDC) and central bank digital account (CBDA). Fan Yifei (2016)[2]argues that the Central bank digital currencies is essentially a pure credit currency that maintains the basic attributes and main features of cash notes. Yao Qian (2016)[3]points out that the Central bank digital currencies are issued by a country's central bank and circulate at the same time as banknotes with the advantages of flexible and convenient transactions and payment-as-settlement, which belongs to the category of cash in circulation. Barrdear & Kumhof (2016)[4]define digital currencies as any form of electronic money characterized by a distributed ledger and a decentralized payment system or an exchange medium. The Bank for International Settlements (2018)[5]released a report on "Central Bank Digital Currency", which puts forward the concept of the "flower of money", which is elaborated from four perspectives, namely, the issuing body is the central bank or other financial institutions, the form of the currency is digitized or physical, whether it is widely available, and whether the technological means is based on the account or based on the tokens. It is proposed that the legal tender digital currency is the center of the "flower of money". Wu Xinhong et al.(2022)[6]argue that legal tender is the Central bank digital currencies issued by the central bank based on cryptography and special transaction algorithms, endorsed by the state's credit, and has the same legal reimbursement as the national credit banknotes.

At present, scholars generally believe that legal tender digital currency will replace part of the cash in circulation and reduce transactional money demand. Zhu Weiliang (2021)[7]establishes the demand function of legal tender by introducing legal tender into the Baumol-Tobin model and conducts a theoretical analysis to conclude that the extent to which the emergence of legal tender reduces the demand for transactional money depends on the transaction cost it saves. Lan Hong (2021)[8]according to Keynes's theory of money demand, respectively, legal digital currency introduced Baumol demand theory to analyze transactional money demand, Tobin model to analyze speculative money demand, Huilun model to analyze precautionary money demand, and carried out theoretical derivation, and came to the conclusion that the circulation of legal digital currency will have a crowding out effect on the demand for money. Zhang Yichao (2019)[9] analyzes the impact of legal digital currency on transactional money demand based on Baumol model theory, and concludes that the emergence of legal digital currency will replace a large portion of cash. Qiu Xun (2017)[10]proposed that in the short and medium term, the People's Bank of China will adopt the simultaneous issuance of digital RMB and banknotes, and the two can be exchanged for each other; and in the long term, it can adopt the model of the parallel existence of digital RMB and cash banknotes and their gradual substitution.Judd & Scaddding (1982)[11]were from the perspective of transactional demand for money, and theoretically investigated how monetary form how innovations in the form of money affect the demand for money and concluded that innovations in the form of money lead to lower switching costs between different assets, which in turn brings about lower public money holdings. Financial analyst Financial analyst Xiao Lei (2020) pointed out to the Daily Economic News reporter that the mission of DC/EP to replace M0 will be gradually realized, and in two to three years' time, 30% to 50% of M0 will be replaced by the central bank's digital currency, and basically realize the nationwide popularization of the central bank's digital currency, and most of the usual cash use will be replaced. This time, the Agricultural Bank of China is the first to promote it, which has a strong practical significance. Pointing out that the mission of digital RMB to replace M0 will be gradually realized, 30% to 50% of M0 will be replaced by the central bank's digital currency in two to three years.

In summary, the current research mainly focuses on the necessity of Central bank digital currencies issuance, issuance mode and operation framework, as well as the impact of Central bank digital currencies on the current monetary policy, macroeconomic operation and RMB internationalization, but the impact of Central bank digital currencieson the demand for transactional currency is less researched and most of the research is still at the theoretical level, lacking a detailed and logical quantitative analysis and demonstration process. However, there are fewer studies on the impact of Central bank digital currencieson the demand for transactional currency and most of them are still at the theoretical level, lacking detailed and logical quantitative analysis and argumentation. The above shortcomings also reserve space for the research of this paper.

### 3 Theoretical analysis

**1.Analysis of the impact of legal tender on transactional money demand**
(1) No statutory digital currency
Baumol-Tobin's theory of money demand builds on Keynesian demand theory by introducing interest rates into the transactional demand for money. He argued that holding cash is an opportunity cost to the public, and that holding cash in the present means foregoing the interest income from investing it. A rational "economic agent" would choose the amount of cash to hold in order to minimize the cost of holding cash.

Assuming that there are two forms of assets in the financial market: interest-bearing assets and cash, the total cost of holding cash for residents is C, Y is the total income in period t and the beginning of the period with all the income held in bonds and other interest-bearing assets, K is the beginning of the period of the cash balance, the end of the period the cash balance of 0, b is the cost of conversion (i.e., the conversion cost of bonds and cash), and r is the interest rate on interest-bearing assets.

The total cost of holding cash $\quad C = \dfrac{K}{2}r + \dfrac{Y}{K}b \quad$ (1)

Assuming that all residents are rational "economic agents", in order to minimize the cost, the equation (1) is obtained by taking the derivative of K and making it zero.

$$\dfrac{\partial C}{\partial K} = \dfrac{r}{2} - \dfrac{bY}{K^2} = 0 \quad (2)$$

$$K = \sqrt{\dfrac{2bY}{r}}$$

In general, people hold, on average, 1/2 of the amount of cash per realization, i.e., K/2, so the money function of the optimal transactional demand $M = \sqrt{\dfrac{bY}{2r}}$ 。

(2).Introduction of legal tender

With the introduction of Central bank digital currencies, residents hold assets in three forms: interest-bearing financial assets, cash, and Central bank digital currencies. In this paper, it is assumed that residents will replace a portion of their cash by converting a portion of their interest-bearing assets into Central bank digital currencies. Assuming that Central bank digital currencies holdings are K and do not generate interest, and that interest on interest-bearing assets is r. The opportunity cost of holding Central bank digital currencies is and the cost of converting interest-bearing financial assets to Central bank digital currencies is $b_1$.

The total cost of holding legal tender digital currency $C_1 = \dfrac{K}{2}r + \dfrac{Y}{K}b_1 \quad$ (3)

Eq. (3) is derived for K to give $\quad \dfrac{\partial C}{\partial K} = \dfrac{r}{2} - \dfrac{b_1 Y}{K^2} = 0 \quad$ (4)

$$K = \sqrt{\dfrac{2b_1 Y}{r}}$$

the optimal transactional money demand function $\quad M_1 = \sqrt{\dfrac{b_1 Y}{2r}} \quad$ (5)

Formally, after the introduction of legal tender in the Baumol-Tobin model, it is still possible to obtain transactional money demand holdings as a square root function. However, it can be seen that the cost of switching financial assets, b1, is reduced because Central bank digital currencies does not incur transaction fees. So it will make M1<M. For transactional money demand, under the Central bank digital currencies condition, a large portion of the transactional demand of residents' cash holdings will be replaced by the Central bank digital currencies, which will have a crowding out effect on the credit cash, thus reducing the transactional money demand.

**2.Mechanisms of impact**

Conversion costs and opportunity costs are two important factors affecting the demand for transactional currencies. Central bank digital currencies affects transactional money demand through two aspects: on the one hand, Central bank digital currencies does not require payment of transaction fees to third-party institutions, and in the future, after banks launch asset management services for Central bank digital currencies, the low conversion cost of Central bank digital currencies provides a convenient channel for asset conversion, and the public can be free from time or space constraints without having to consider the conversion cost, so that Central bank digital currenciescan be converted without the need to consider the conversion cost by accelerating the speed of inter-conversion between interest-bearing assets and transactional currency demand, improving the convenience of conversion, accelerating conversion between currency tiers, and reducing the costs and fees of conversion between currencies of different tiers in a way that will continue to have a substitution effect on the demand for physical RMB. On the other hand, the development of legal tender digital currency payments has given rise to a trend of gradual monetization of financial assets, i.e., interest-bearing assets with high yields and high liquidity at the same time. Therefore, the public can quickly convert interest-bearing assets into transactional currencies to satisfy their daily payment needs without the need to hold more transactional currencies, which reduces the opportunity cost of holding transactional currencies. As a result, the public will choose to hold more high-yield financial assets and hold less transactional currency demand. In summary, the issuance and circulation of Central bank digital currencies will have a substitution effect on cash through the above two effects, blurring the clear boundaries between money demands and reducing the amount of transactional money demand.

## 4 Model construction and data selection

### 1.Model construction

(1) Semi-logarithmic model construction

Based on the research purpose of this paper, the semi-logarithmic model of transactional money demand is constructed as follows:

$$\ln M0 = c_0 + c_1 MDI + c_2 A + c_3 r + c_4 \ln GDP + \varepsilon_t \qquad (6)$$

$\varepsilon_t$ represents the residual term, $c_1$、$c_2$、$c_3$、$c_4$ represents the coefficient of each explanatory variable.

(2) Autoregressive Distributed Lag Model ARDL Model Construction

For the false regression caused by the non-stationary characteristics of time series data, this paper adopts the cointegration test for long-term stability test. This paper adopts the ARDL Bounds Testing method (Bounds Testing) proposed by Pesaran (2001)[13]and other scholars, which has obvious advantages. Because the sample size of quarterly data for 2010-2022 selected in this paper is 52, the ARDL model is suitable for empirical analysis. The model construction is divided into the following three steps:

The first step is to construct the ARDL Boundary Cointegration Test model:.

$$\Delta \ln M0 = a_1 + \sum_{i=1}^{p1}\beta_{1i}\Delta \ln M0_{t-i} + \sum_{i=1}^{p2}\beta_{2i}\Delta MDI_{t-i} + \sum_{i=1}^{p3}\beta_{3i}\Delta A_{t-i} + \sum_{i=1}^{p4}\beta_{4i}\Delta r_{t-i}$$
$$+ \sum_{i=1}^{p5}\beta_{5i}\Delta \ln GDP_{t-i} + \mu_1 \ln M0_{t-1} + \mu_2 MDI_{t-1} + \mu_3 A_{t-1} + \mu_4 r_{t-1} + \mu_5 \ln GDP_{t-1} + \varepsilon_{1t} \quad (7)$$

$\Delta$ denotes first-order differencing of the variables; $a_1$ is a constant term; $p_i(i=1,2,3,4,5)$ is the optimal lag order, which is chosen in this paper to be determined by the SBC criterion; $\mu_i(i=1,2,3,4,5)$ denotes the long-term correlation coefficient of the model; $\beta_{ki}(k=1,2,3,4,5)$ is the short-term dynamic correlation coefficient of the model; $\varepsilon_{1t}$ is a white noise sequence obeying a normal distribution.

The second step, the following ARDL model is developed to estimate the long-run coefficients.

$$\ln M0_t = a_2 + \sum_{i=1}^{\delta 1}\eta_{1i}\ln M0_{t-i} + \sum_{i=1}^{\delta 2}\eta_{2i}MDI_{t-i} + \sum_{i=1}^{\delta 3}\eta_{3i}A_{t-i} + \sum_{i=1}^{\delta 4}\eta_{4i}r_{t-i} +$$
$$\sum_{i=1}^{\delta 5}\eta_{5i}\ln GDP_{t-i} + \varepsilon_{2t} \quad (8)$$

The third step, the short-term dynamic coefficients are derived from the ARDL error correction (ARDL-ECM) model by.

$$\Delta \ln M0_t = a_3 + \sum_{i=1}^{\sigma 1}\gamma_{1i}\Delta \ln M0_{t-i} + \sum_{i=1}^{\sigma 2}\gamma_{2i}\Delta MDI_{t-i} + \sum_{i=1}^{\sigma 3}\gamma_{3i}\Delta A_{t-i} + \sum_{i=1}^{\sigma 4}\gamma_{4i}\Delta r_{t-i} +$$
$$\sum_{i=1}^{\sigma 5}\gamma_{5i}\Delta \ln GDP_{t-i} + \lambda ECM_{t-i} + \varepsilon_{3t} \quad (9)$$

$ECM_{t-1}$ is the error correction term, and the error correction factor is given by $\lambda ECM_{t-1}$, its coefficient represents the self-correcting speed $0<\lambda<1$.

### 2. Variable selection

(1) Explained variable transactional money demand: Referring to Chen Tiefei (2006) [13] that the money market is always in ex-post equilibrium, i.e., the demand for money is comparable to the supply of money, and the demand for money can be measured by the supply of money. So in this paper, quarterly M0 is chosen to represent the transactional money demand, and the source of data is the People's Bank of China.

(2) Core Explanatory Variable Central bank digital currencies: as the digital RMB is still in the pilot stage, it has not been formally launched to the public, and the specific issuance data is unknown. At present, China's Central bank digital currencies is mainly a substitute for the transfer of M0 and third-party mobile payments, and Central bank digital currencies is currently mainly traded through the binding of bank cards, according to Shi Xinlu et al. (2018)[14]. The measurement method and data processing of third-party payment as well as electronic money, comprehensive consideration of China's future issuance of D/EP to third-party payment and credit cash is a gradual replacement. So this paper constructs the MDI instead of the future issuance of Central bank digital currencies in the future, where MDI = (online payment amount + bank card transfer amount + bank card consumption balance) / GDP, the size of the MDI to a certain extent also represents the breadth of the Central bank digital currencies issuance to be used in the future. Source of data: The People's Bank of China (PBOC) quarterly "Payment System Operation General Situation" and Wind database.

(3)Gross output (GDP): quarterly data of gross national product (current price) is selected, and the source of data is the National Bureau of Statistics.

(4)Conversion cost (A): Due to data availability, this paper selects the commission when converting stocks to credit cash as the conversion cost, which mainly includes stamp duty and brokerage commission rate. Where brokerage commission rate = brokerage commission / stock trading volume, select the stock trading volume of Shanghai and Shenzhen, data source Wind database. The semi-annual data source of brokerage commission is the database of fund trading database, and the stamp duty rate has been kept at 0.1% since 2008.seat rental and trading status of Rexis .

(5)Opportunity cost (r): Normally, the public meets its transactional money needs by holding cash, but this approach results in the loss of interest generated by holding other interest-bearing financial assets, increasing the opportunity cost. With the deepening of China's market-oriented interest rate reform, in order to better reflect market supply and demand, this paper selects the monthly data of the 7-day interbank lending weighted rate to reflect the opportunity cost. Data source Wind database.

**3.data processing**

First, the missing brokerage commission data are supplemented with the help of SPSS software. Second, this paper adjusts the semiannual data of brokerage commissions to quarterly data by total interpolation method with the help of Eviews 10 software, and converts the monthly data M0, monthly data stock turnover and monthly data 7-day interbank lending weighted rate to quarterly data. Since time series observations often have cyclical movements, all data are seasonally processed using the Census X-12 method. Finally, to avoid the existence of multicollinearity and heteroskedasticity, the seasonally processed GDP and aggregate transactional money demand data are taken to be natural logarithmized in this paper and denoted as LNGDP and LNM0, respectively.The descriptive statistics of each variable are shown in Table 1.

Table 1 Variable definitions and descriptive statistics

|  | LNGDP | LNM0 | A | MDI | r |
| --- | --- | --- | --- | --- | --- |
| average value | 12.1327 | 11.0827 | 0.1017 | 7.7092 | 3.0787 |
| median | 12.1323 | 11.0908 | 0.1013 | 8.7715 | 3.1494 |
| Maximum value | 12.6267 | 11.5482 | 0.1044 | 11.2676 | 4.6973 |
| Minimum value | 11.4774 | 10.5420 | 0.0997 | 3.5044 | 1.6471 |
| Standard deviation | 0.32755 | 0.2442 | 0.0010 | 2.7615 | 0.7567 |
| Skewness | -0.1955 | -0.1908 | 1.0245 | -0.2365 | 0.3057 |
| Kurtosis | 1.9846 | 2.4220 | 3.8044 | 1.3565 | 2.4024 |
| JB normality test | 2.5650 | 1.0394 | 10.499 | 6.3367 | 1.5837 |
| P-value | 0.2773 | 0.5947 | 0.0052 | 0.0421 | 0.4530 |

Source: Wind database and rhe People's Bank of China.

## 5 Analysis of empirical results

### 1. Smoothness test and cointegration test for variables

(1) ADF test

The ADF test for each variable is given in Table 2, the original series of transactional money demand (LNM0), aggregate output (LNGDP) and opportunity cost r are smooth at the 5% level and are zero-order single integer series. The legal tender digital currency (MDI) and conversion cost A are first order single integer series.

Table 2 Smoothness test for each variable

| Variables | Test form (C, T, L) | t-value | P-value | Test result |
|---|---|---|---|---|
| LNM0 | (C,T,0) | -3.5029 | 0.0497 | steady |
| MDI | (C,0,0) | -1.5433 | 0.5039 | non-stationary |
| DMDI | (C,T,0) | -6.1493 | 0.0000 | steady |
| A | (C,T,8) | -2.4097 | 0.3695 | non-stationary |
| DA | (C,0,7) | -3.5305 | 0.0118 | steady |
| r | (C,T,0) | -3.7146 | 0.0302 | steady |
| LNGDP | (C,T,0) | -3.8671 | 0.0208 | steady |

### 2. Marginal cointegration test

The unit root test shows that all variables are consistent with I(1) and I(0) and do not exceed I(1), allowing the ARDL marginal cointegration test to be used. The original hypothesis of equation (8) in this paper is $H_0: \mu_1 = \mu_2 = \mu_3 = \mu_4 = 0$, alternative hypothesis $H_1: \mu_1 \neq 0$ or $\mu_2 \neq 0$ or $\mu_3 \neq 0$ or $\mu_4 \neq 0$。Referring to the critical values of upper and lower bounds given by Pesaran et al. (2001) [17], I(1) is the upper limit of the critical value, and I(0) is the lower limit of the critical value, the original hypothesis is rejected when the overall significant F-statistic is greater than I(1); the original hypothesis is accepted if F-statistic is less than I(0); and the boundary cointegration test fails if the F-statistic is between the two critical values. The results of the borderline cointegration test are shown in Table 3, the F statistic is 8.53, which is greater than the 1% critical value of 4.37, and it is considered that there is a long-run equilibrium relationship between the variables and the cointegration relationship is significant.

Table 3 Boundary cointegration test result

| Significance Level | 10% | | 5% | | 1% | |
|---|---|---|---|---|---|---|
| | I(0) | I(1) | I(0) | I(1) | I(0) | I(1) |

| Critical value | 2.2 | 3.09 | 2.56 | 3.49 | 3.29 | 4.37 |
| --- | --- | --- | --- | --- | --- | --- |
| F-statistic | F<LNM0 \| MDI,A,r,LNGDP>=8.5299 | | | | Cointegration relationship | |

**3.ARDL model estimation and description**

In this paper, we use Mircofit5.5 software to firstly estimate the ARDL model long-run equilibrium coefficients and further run the ARDL-ECM model to estimate the short-run model coefficients. In this paper, the SBC criterion is chosen to select the optimal lag order of (1,0,0,1,0), and the model estimation results are shown in Table 4.

In both the long-run and the short-run, legal tender, opportunity cost and total output have significant effects on the demand for transactional money. The coefficients on legal tender and opportunity cost are significantly negative and the coefficient on total output is significantly positive. The conversion cost is not significant probably because the data selected in this paper is only the conversion cost of stocks, without considering financial assets such as bonds and funds, which is not comprehensive enough.

The long-term and short-term coefficients of transactional money demand and legal tender are both significantly negatively correlated at the 1% level, and the long-term coefficient is -0.026702, indicating that legal tender will reduce transactional money demand by 0.027% when legal tender is raised by 1%; and the impact in the short term is smaller. The empirical results show that in the short term, the substitution effect of legal tender on transactional currency is small, but in the long term, legal tender will affect the structure of China's currency demand and reduce transactional currency demand by virtue of its convenience, security and other advantages, which is in line with the theoretical model derivation.Both the long-term coefficients and the short-term coefficients between the LNGDP and transactional currency demand are significantly positive at the 1% level. correlation, indicating that an increase in GDP causes an increase in consumption, income and investment, and more living expenses of residents, which ultimately causes a significant increase in residents' money demand.GDP is one of the most important influences on money demand in the long run. Transactional money demand is more responsive to opportunity cost, significant at the 1% level in the short run. As the circulation of legal tender gradually promotes the securitization of financial markets, the proportion of cash and deposits in the public's asset portfolios will decline and the proportion of interest-bearing financial assets will increase, suggesting that the opportunity cost of holding transactional money demand has an enhanced impact on the public's asset allocation behavior. As the legal tender will make transactional money demand more sensitive and elastic to financial market interest rates, this will increase the difficulty of the central bank's monetary policy implementation, change the structure of China's money demand, and reduce the effectiveness of the central bank's quantitative monetary policy. Conversion cost and transactional money demand in the long and short term are positively correlated, indicating that the lower the conversion cost of legal tender and other financial assets, the higher the frequency of conversion of various types of assets, the lower the transactional money demand, which will reduce the stability of China's money demand. The error correction term ECM is significant at the 1% level with a coefficient of -0.22703, implying that when transactional money demand is affected in the short term, it will be adjusted to the long-run equilibrium value in the next year at a rate of 22.7% in real terms, which also indicates the appropriateness of the constructed ARDL-ECM model. The model diagnostic section presents the regression results with an R2 of 0.99560, which indicates a good model fit; the LM test shows that there is no autocorrelation, and the Ramsey test indicates that the model does not omit key variables and is set correctly.

Table 4  ARDL Model Long and Short-Term Coefficient Estimates

|  | Dependent variable | Coefficient | Standard error | t-statistic[P-value] |
|---|---|---|---|---|
| Estimation of long-term coefficients | MDI*** | -0.0267 | 0.0092 | -2.8964 [0.006] |
|  | A | 2.2097 | 15.6598 | 0.1411 [0.888] |
|  | $r^*$ | -0.0457 | 0.0229 | -1.9875 [0.053] |
|  | LNGDP*** | 0.8361 | 0.0737 | 11.3526 [0.000] |
|  | INPT | 1.1487 | 2.0572 | 0.5584 [0.579] |
| Short-term ECM model estimation results | $\Delta MDI$*** | -0.0061 | 0.0022 | -2.8011 [0.007] |
|  | $\Delta A$ | 0.5017 | 3.4848 | 0.1440 [0.886] |
|  | $\Delta r$*** | -0.0104 | 0.0035 | -2.9666 [0.005] |
|  | $\Delta$LNGDP*** | 0.0068 | -0.0892 | -2.8011 [0.007] |
|  | ECM(-1)*** | -0.2270 | 0.0849 | -2.6732 [0.01] |
| Model Diagnostics | | | | |
| LM | DW | $R^2$ | Adjustment $R^2$ | Ramsey test[P 值] |
| 6.6648 [0.155] | 2.2678 | 0.99613 | 0.9956 | 0.154 [0.8784] |

注：* ** ***represent significant at 10%, 5%, and 1% significance levels, respectively。

$d \ln M0 = \ln M0 - \ln M0(-1)$、 $dA = A - A(-1)$  $dMDI = MDI - MDI(-1)$、 $dr = r - r(-1)$、

$d \ln GDP = \ln GDP - \ln GDP(-1)$ 、INPT is the intercept term,

$ECM = \ln M0 + 0.0267 MDI - 2.2097 A + 0.0457 r - 0.8361 \ln GDP$

## 4. Stability Test

Pesaran et al. (2001) [17] recommended the use of CUSUM and CUSUMSQ to test the stability of the residuals of the error correction equation. As shown in the figure, the fluctuation range of CUSUM and CUSUMSQ values are within the confidence band of 5% significant level, so the regression parameter estimates are stable and plausible in terms of the overall stability of the model.

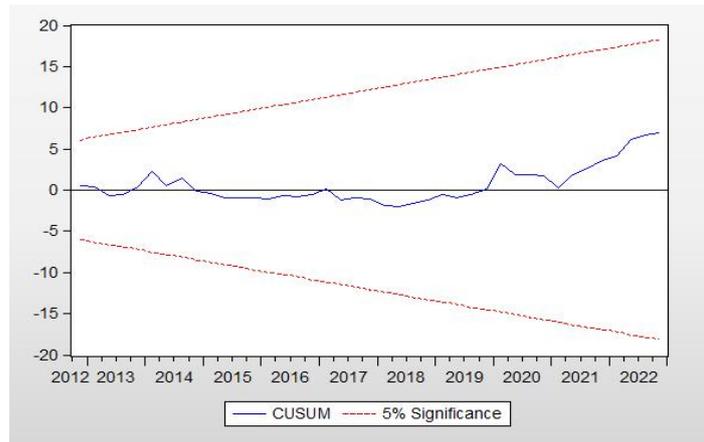

Fig 1    Recursive residual accumulation and CUSUM

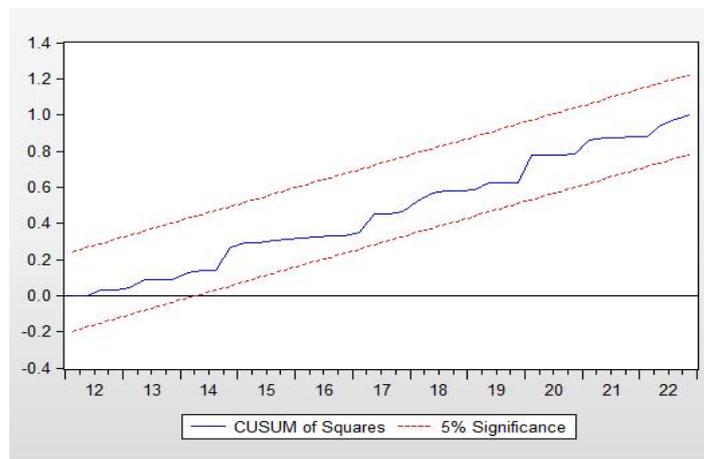

Fig 2    Recursive residual squared accumulation and CUSUMSQ

## 6 Conclusion and Recommendations

This paper theoretically analyzes the impact of legal digital currency on the demand for transactional money, and finds through theoretical analysis that legal digital currency will reduce the demand for transactional money. And the quarterly time series data from 2010 to 2022 are selected, and the ARDL model is established for empirical analysis. The empirical results show that there is a significant negative correlation between legal tender and transactional money demand, and legal tender will significantly reduce transactional money demand in the long and short term, leading to the instability of the structure of money demand, which will increase the difficulty of the central bank in forecasting and grasping the total amount of money demand. Based on the above conclusions, relevant suggestions are put forward in accordance with the current situation of the development of legal tender in China:

First, the central bank should continue to strengthen the underlying technical support foundation of the Central bank digital currencies. First, it is necessary to strengthen the research and development and innovation of blockchain technology in the research and development and management of China's Central bank digital currencies, improve the digital currency technical standard system, continuously optimize the design of the digital renminbi system, maintain the overall technological sophistication, and provide a solid bottom-up support for the innovation of the Central bank digital currencies business and application, and then guarantee that the digital currency system can stably realize the characteristics of the Central bank digital currencies, and ensure that it can be safe and stable circulation; continue to improve the security management of its operation system, and provide a perfect technology and service system for the popularization of Central bank digital currencies. Second, accelerating the construction of a digital central bank will help improve the timeliness and precision of macroeconomic policies and realize real-time sharing of financial market information.

Secondly, the construction of application scenarios should be strengthened, and the popularization of Central bank digital currencies should be actively promoted. Although China's Central bank digital currencies is accelerating its entry into the homes of ordinary people, the public's awareness and acceptance of Central bank digital currencies as a novelty needs to be improved. In terms of application scenarios, it is necessary to gradually form a full range of application scenarios for Central bank digital currencies in daily consumption, payment, transportation, education and medical care, culture, sports and tourism. In the promotion process, emphasis has been placed on simplifying the payment process, realizing diversified payment methods for different groups of people, and including the elderly population in the scope of application of Central bank digital currencies to narrow the digital divide.

Thirdly, attention is paid to the changes in the demand for money by micro-entities. The issuance of legal tender digital currencies will affect the relationship between the original variables in money demand, leading to instability in the forecast of money demand. Changes in money demand will exacerbate the difficulty of matching money supply with money demand, and will reduce the effectiveness of the implementation of China's quantitative monetary policy with money supply as the intermediary target. Therefore, it is recommended that the Central Bank make legal tender a highly efficient monetary policy tool after its issuance through reasonable mechanism settings, giving full play to the advantages of legal tender and making monetary policy transmission more precise.

Fourth, strengthen the legislative framework and policies supporting the industry in relation to Central bank digital currencies. To control the development of Central bank digital currencies, provide safeguards for their long-term benign development from the system, consolidate the rule of law and regulatory basis for their development in China, supplement China's Regulations on the Administration of the Renminbi, and expand the contents of Central bank digital currencies, it is necessary for the Central Bank and relevant departments to develop laws and regulations.In order to create a smooth and sensitive monetary policy interest rate transmission mechanism and ultimately achieve the monetary policy objectives through the regulation of market interest, the central bank should enhance the market benchmark interest rate system with Shibor as the core. It should also gradually switch the monetary policy control mode to one that is interest rate-oriented. The central bank can take use of the novelty of monetary policy instruments provided by the advent of Central bank digital currencies to better fulfill the purpose of monetary policy.

## [References]

[1] Koning J.P.,2017.Evolution in cash and payments:comparing old and new ways of designing central bank payments systems,cross-border payments networks,and remittances.R3 Reports.

[2] Fan Yifei. Theoretical basis and architectural choice of China's legal tender digital currency[J]. China Finance,2016(17):10-12.

[3] Yao Qian. China's legal digital currency prototype conception[J]. China finance,2016(17):13-15.

[4] Barrdear&Kumhof. The Macroeconomics of Central Bank Issued Digital Currencies. Staff Working Paper, Bank of England, 2016.

[5] Bank for International Settlements. "Central bank digital currencies, BIS Committee on Payments and Market Infrastructures". Market Committee

[6] Wu Xinhong,Pei Ping. Legal tender digital currency: theoretical foundation, operation mechanism and policy effect[J]. Journal of Soochow University (Philosophy and Social Science Edition),2022,43(02):104-114.

[7] Zhu Weiliang,Dong Chao,Cai Ran. Central Bank Digital Currency and the Relationship between Currency Supply and Demand[J]. Financial Market Research,2021(10):39-49.

[8] Lan Hong,Yang Wen,Wei Dongyun. The Impact of Legal Digital Currency on China's Structural Monetary Policy[J]. Southwest Finance,2021(11):89-100.

[9] Zhang Yichao,Xu Guocheng. Exploration of the impact of legal digital currency on the demand and supply of money[J]. North Finance,2019(03):43-47.

[10] Qiu Xun. The issuance of digital currency by China's central bank:path, problems and its response strategy[J]. Southwest Finance,2017(03):14-20.

[11] Judd J. P.,Scadding J. L. The Search for a Stable Money Demand Function: A Survey of the Post-1973 Literature[J]. Journal of Economic Literature, 1982, 20(3): 993-1023.

[12] Pesaran M H, Shin Y, Smith R J. Bounds testing approaches tothe analysis of the level relationships [J]. Journal of AppliedEconometrics, 2001,16(3):289-326.

[13] Chen Tiefei. Validation of China's money demand model based on financial innovation factors[J]. Shanghai Finance,2006(03):32-35.

[14] Shi Xinlu, Zhou Zhengning. Electronic payment, currency substitution and money supply[J]. Research in Financial Economics,2018,33(04):24-34.